# Comments on "All-electron mixed basis *GW* calculations of TiO2 and ZnO crystals"

Diola Bagayoko,[1] Yacouba Issa Diakité,[2] Chinedu E. Ekuma,[3] Lashounda Franklin[1]

[1]Department of Mathematics and Physics, Southern University and A&M College
in Baton Rouge (SUBR), Baton Rouge, LA 70813, USA
[2]Faculté des Sciences et Techniques (FST), Université des Sciences,
des Techniques, et des Technologies de Bamako (USTTB), Bamako, Mali
[3]NRC Research Associate, Theoretical Chemistry, Naval Research Laboratory (NRL), 4555
Overlook Ave., SW Washington, D.C. 20375, USA

These brief comments on the article in Phys. Rev. B 93, 155116 (2016), address an inadvertent misrepresentation of the capabilities of density functional theory (DFT) and of its local density approximation (LDA) in describing electronic and related properties of materials accurately. The oversight of some previous LDA results that agree with experiments partly led to this unintended misrepresentation, in addition to a few assertions relative to perceived deficiencies of DFT.

Zhang et al[1] recently published the results from their mixed basis set Green Function and dressed Coulomb (GW) approximation calculations of electronic properties of anatase and rutile $TiO_2$ and of zinc blende and wurtzite ZnO. They employed plane waves and atomic orbitals to better describe the empty and bound states, respectively. They utilized a full $\omega$ integration to evaluate the correlation part of the self-energy and introduced a new approach for this integration. They equally employed plasmon pole models and the generalized plasmon pole model (GGP) for the referenced evaluation. Zhang *et al.*,[1] performed single shot GW calculations that are often referred to as $G_oW_0$ approximation. These calculations are non-self-consistent, even though their results, for the band gaps, are generally much better than those from the single basis set density functional theory (DFT) and local density approximation (LDA) computations. This significant improvement does not, however, correct for violations, by $G_0W_0$ calculations, of momentum, energy, and particle conservation laws.[2]

The findings of Zhang *et al.*, for the band gaps of the noted structures of $TiO_2$ and of ZnO, are summarized in Table I below, along with those from several, previous GW calculations.[3-17] Zhang and coworkers thoroughly discussed these previous results in comparison to theirs. As per the contents of Column 2, their LDA calculations woefully underestimated the band gaps of $TiO_2$ and of ZnO. There is no indication that these LDA calculations employed successively larger basis sets, beginning with a relatively small one, to search for and to attain the absolute minima of the occupied energies – as required by the second Hohenberg-Kohn theorem also known as the DFT variational principle. In contrast to the LDA results of Zhang and coworkers, the ones (in Column 8) obtained with the Bagayoko, Zhao, and Williams (BZW)[20-21] method, as enhanced by Ekuma and Franklin (BZW-EF),[18-19,22] are in excellent agreement with the corresponding, experimental ones for both rutile[18] $TiO_2$ and wurtzite[19] ZnO. Consequently, the oversight of these results by Zhang *et al.*[1] leads to an inadvertent misrepresentation of the capabilities of DFT and LDA in describing electronic properties of materials, as further explained below.

**Table I**. Excerpts from Table I of Zhang *et al*. Band gaps of rutile TiO2 and wurtzite ZnO in electron volts (eV). Columns 2-5 show the LDA and GW results of Zhang *et al*. Previous GW results are in Column 6. For the previous GW results, values without explanation in parentheses denote the one-shot *GW* method, while sc*GW* denotes the self-consistent *GW* method, and so on. GPP, one-shot *GW* method using the GPP model; vLH, plasmon pole model of von der Linden–Horsch; $\omega$'-int., $\omega$' integration. Ind. and Dir. stand for indirect and direct.

| | LDA[1] | GW Results of Zhang *et al.*[1] | | | Previous GW | Expt. | BZW & BZW-EF |
|---|---|---|---|---|---|---|---|
| | | GPP | vLH | $\omega$'-int. | | | |
| Rutile TiO$_2$ | 1.76 | 4.0 | 3.3 | 3.30 | 4.48(GPP),[a] 3.34,[a] 4.8,[b] 3.59,[c] 3.30,[d] 3.78(sc*GW*),[e] 4.48(sc*GW* with GGA + *U*)[f] | 3.3 ± 0.5[g] | 2.95[p] Ind. 3.05[p] Dir. |
| Wurtzite ZnO | 1.10 | 4.54 | 3.91 | 2.82 | 2.44,[h] 4.23(model *GW*),[i] 3.4(GPP),[j] 3.6(GPP/LDA + *U*),[j] 2.83,[k] 2.352,[l] 2.56,[m] 3.88(*GW$^l$*)[n] | 3.4[o] | 3.39[q] Dir. |

[a]Reference [3].
[b]Reference [4].
[c]Reference [5].
[d]Reference [6].
[e]Reference [7].
[f]Reference [8].
[g]Reference [9].
[h]Reference [10].
[i]Reference [11].
[j]Reference [12].
[k]Reference [13].
[l]Reference [14].
[m]Reference [15].
[n]Reference [16].
[o]Reference [17].
[p]Reference [18].
[q]Reference [19].

Bagayoko[22] presented an understanding of DFT that adheres to conditions inherent to its validity. Specifically, Bagayoko underscored the fact that single basis set calculations using DFT (including LDA) potentials lead to stationary solutions that are potentially infinite in number. Such solutions do not generally describe the ground state of the system under study. One reaches such stationary solutions with any reasonable choice of the basis set. The second Hohenberg theorem, however, clearly states that the ground state is reached only if one employs the "correct" ground state charge density, which is *à priori* unknown. The BZW or BZW-EF method rests in part on the fact that, alternatively, if one searches and finds the absolute minima of the occupied energies (i.e., representing the ground state), then the corresponding, self-consistent charge density should be that of the ground state. As noted by Bagayoko et al.[20] and by Zhao and coworkers,[21] as early as 1998 and 1999, respectively, one should not ignore the fact that this charge density is calculated using the wave functions of the occupied states only.

Indeed, the use of only the wave functions of the occupied states points to the absence of a one-to-one correspondence between the self-consistent charge density and the basis set. As shown in our studies of tens of semiconductors,[22] there is potentially an infinite number of basis sets that lead to the same self-consistent ground state density. The smallest of these basis sets, known as the optimal basis set,[18-19,20-22] is the *only* one for which both the occupied and low, unoccupied (up to 10 eV) energies can be ascribed to DFT. In the successive augmentation of the basis set, starting with a small one, both occupied and unoccupied energies are generally lowered from one calculation to the next, up to the attainment of the absolute minima of the occupied energies. Once these minima, i.e., the ground state, are reached, with the optimal basis set, the occupied energies no longer change following an augmentation of the basis set. There exists an infinite number of such large basis sets that include the entirety of the optimal one. Several calculations discussed by Bagayoko[22] show some of these basis sets larger than the corresponding, optimal ones for zinc blende ZnS and BP and for wurtzite GaN and ZnO. Due to the aforementioned asymmetry between wave functions of occupied and unoccupied states, these basis sets lead to the same charge density, Hamiltonian content, and occupied energies. However, for basis sets much larger than the optimal one, the Rayleigh theorem leads to an artificial lowering of some unoccupied energies from their values obtained with the optimal basis set. This extra lowering, cannot be attributed to any interaction in the Hamiltonian that does not change from its value obtained with the optimal basis set.

The Rayleigh theorem states that when the *same* eigenvalue equation is solved with N and (N+1) functions, where all the N functions are included among the (N+1), and the resulting eigenvalues are ordered from the lowest to the highest, then any one of the first N eigenvalues from the second calculation is lower or equal to the corresponding one from the first calculation with N functions. [The reader should note that "the *same*" eigenvalue equation does not just mean two equations of the same form, but also with the same content, numerical and otherwise, for all the operators involved in the two equations.] *Once the absolute minima of the occupied energies are reached, larger basis sets that include the optimal one lead to the extra lowering of some unoccupied energies, including some lowest ones, while the occupied ones do not change.* By virtue of the first DFT theorem, the spectrum of the Hamiltonian and its energy content (here the sum of occupied energies) are unique functional of the ground state charge density! As explained by Bagayoko,[22-23] while occupied and low, unoccupied energies obtained with the optimal basis set belong to the spectrum of the Hamiltonian, the unoccupied energies that are lowered from their value obtained with the optimal basis set no longer belong to the spectrum of the Hamiltonian that did not change from its value for the optimal basis set.

Incidentally, the basis set size dependence of the extra lowering of unoccupied energies not only explains in part the recalcitrant underestimation of band gaps by single basis set DFT and LDA calculations, but also elucidates the fact that several of these calculations disagree with each other even if they employ the same DFT potential and the same computational approach. They only need to have basis sets of different sizes (and possibly other features), which hopefully contain the optimal one, in order to have different, unoccupied energies and different band gaps. With the inclusion of the optimal one in these basis sets, the occupied energies, including the top of the valence bands, possess the complete, physical content of DFT. *However, extra-lowering of unoccupied energies, including the bottom of the conduction band, destroyed the full, physical content they had when obtained with the optimal basis set.*

The above extra-lowering of some unoccupied energies, with large basis sets that contain the optimal one, can be unwittingly misinterpreted as the proof that the unoccupied energies did not converge with the optimal basis set. Such a misinterpretation is tantamount to expecting to obtain new, unoccupied energies, i.e., excited state energies, from a Hamiltonian that does not change in content: once the optimal basis set is reached, the charge density no longer changes; so, the cannot change. This fact should suffice for understanding the destruction of the DFT or physical content of unoccupied energies with the use of basis sets larger than the optimal one and that contain it. Another way of grasping this fact is that, by virtue of the Rayleigh theorem, there is no known lower bound to the extra-lowering. So, the resulting, erroneous band gap from calculations with these larger basis sets can, for instance, take any value between zero and the correct band gap- depending on the sizes of the basis sets.

The above discussions pertained to the inadvertent misrepresentation of the capabilities of DFT to describe electronic properties of materials. Specifically, they explain the reasons that single-basis set DFT calculations cannot be expected to produce the correct description of materials. Additionally, they spelled out the physics and mathematical reasons (i.e., the two Hohenberg-Kohn and the Rayleigh theorems) that the BZW-EF method leads to numerical results that possess the full, physical content of DFT. In light of these points, the assertions of Zhang *et al.* relative to the deficiencies of Kohn-Sham (KS) eigenvalues in representing quasi-particle energies should be circumscribed (i.e., qualified). On the one hand, from the above discussions, the assertions are almost universally true for KS eigenvalues obtained from single basis set calculations; on the other hand, they do not hold for KS eigenvalues obtained from calculations that rigorously follow the BZW-EF method. The above circumscription also applies to the assertion by Zhang *et al.*, in the presentation of their results, that the band structures, in LDA calculations, are not improved by local and energy-independent exchange correlations. While this statement is correct for single basis set LDA calculations, it is not for those employing the BZW-EF method.

In closing, we hope to have shown, with LDA results in accord with experiment, that the oversight of these results inadvertently leads to a misrepresentation of the capabilities of DFT and LDA in describing electronic and related properties of materials. In doing so, we hope to have demonstrated the reasons that single basis set DFT calculations lead to stationary solutions that cannot be expected, by chance, to be ground state solutions. The attainment of the latter requires the search for and the attainment of the optimal basis set. Additionally, DFT and LDA calculations have to avoid the destruction of the physical content of the low, unoccupied energies with the use of basis sets much larger than the optimal one and that contain it. Each of the Rayleigh theorem and the first Hohenberg-Kohn suffices to do so, on mathematical and physical grounds. These large basis sets, over-complete for the description of the ground state, can produce any band gap value between zero and the actual, measured one.

**Acknowledgments:** This work was funded in part by the US Department of Energy, National Nuclear Security Administration (NNSA, Award Nos. DE-NA0001861 and